\documentclass[preprint,prb,amsmath,amssymb,amsfonts,superscriptaddress,floatfix,showpacs]{revtex4}
\usepackage{graphicx}% Include figure files
\usepackage[dvips]{color}
\usepackage{dcolumn}% Align table columns on decimal point
\usepackage{bm}% bold math
\usepackage{color}

\begin{document}

\title{Groundstate electronic structure of actinide carbides and nitrides}

\author{L. Petit}
 \email{lpetit@phys.au.dk}
 \affiliation{Department of Physics and Astronomy, University of Aarhus, DK-8000 Aarhus C, Denmark}
\author{A. Svane}
 \affiliation{Department of Physics and Astronomy, University of Aarhus, DK-8000 Aarhus C, Denmark}
\author{Z. Szotek}
 \affiliation{ Daresbury Laboratory, Daresbury, Warrington WA4 4AD, UK }
\author{W. M. Temmerman}
 \affiliation{ Daresbury Laboratory, Daresbury, Warrington WA4 4AD, UK }
\author{G. M. Stocks}
 \affiliation{ Materials Science and Technology Division, Oak Ridge National Laboratory, Oak Ridge, Tennessee 37831, USA }

\date{\today}
\begin{abstract}
The self-interaction corrected (SIC) local spin-density approximation (LSD) is used to investigate the groundstate valency
configuration of the actinide ions in the actinide mono-carbides, AC (A = U, Np, Pu, Am, Cm), and the actinide mono-nitrides, AN.
The electronic structure is characterized by a gradually increasing degree of $f$-electron localization from U to Cm, with the
tendency towards localization being slightly stronger in the (more ionic) nitrides compared to the (more covalent) carbides.
The itinerant band-picture is found to be adequate for UC and acceptable for UN, whilst a more complex manifold of competing
localized and delocalized $f$-electron configurations
underlies the groundstates of NpC, PuC, AmC, NpN, and PuN. The fully localized 5$f$-electron configuration is realized in
CmC ($f^7$), CmN ($f^7$), and AmN ($f^6$).
The observed sudden increase in lattice parameter from PuN to AmN is found to be related to the localization transition.
The calculated valence electron densities of states are in good agreement with photoemission data.

\end{abstract}

\maketitle

\section{Introduction}
Among the six generation IV reactor technologies that have been chosen to address the future nulear energy requirements,
three are so-called fast neutron reactors.~\cite{worldnuclear}
Given their efficient use of existing Uranium-ore (fast breeders) and the possibility of burning the higher actinides,
i.e. disposing of the nuclear
waste produced by existing thermal nuclear reactors, these reactors address a number of concerns surrounding the issue of nuclear energy.
As there are no moderators, the fission reactions depend on fast neutrons, requiring a small core with a high power density and
very efficient heat transfer.
The ongoing research and development is therefore
considering possible alternatives to the predominantly oxide based fuels.
Both carbide and nitride fuels are being investigated for this purpose
given their superior thermophysical properties ~\cite{Srivastava} such as, high melting point, high heavy atom density, and high thermal
conductivity, and, with respect to Na cooled fast reactors, good compatibility with the coolant.
Compared to the well established oxide fuels, relatively few studies exist regarding the physics and chemistry of these materials. Thus,
modeling of the structural and dynamical properties under ambient or operating conditions can provide valuable information
concerning fuel performance and stability.
In this paper the focus is on groundstate properties where we use density functional based total
energy calculations to study the electronic structure of the actinide mono-carbides and mono-nitrides.

Due to the onset of 5$f$-electron localization phenomena, the theoretical description of the actinide compounds
presents a considerable challenge.
While band structure calculations, based on the local spin-density approximation (LSDA) to density functional theory, are very successful in describing the cohesive properties of itinerant electron systems they have serious problems when dealing with more strongly correlated electron systems. The reason for this is that the exchange and correlation effects underpinning the standard 
LSDA approaches are those of the homogeneous electron gas, which cannot
account for the strong electron-electron interactions that are inherent to $f$-electron sytems.

In the actinide metals for example, the localization transition that occurs from Pu to Am is not correctly reproduced.~\cite{Johansson_AcMetal}
In the early actinide metals the overlap between $f$-orbitals on neighbouring sites results in $f$-electron delocalization
and band formation. However, with increasing nuclear charge the $f$-orbitals contract with the result that, in the late actinides, the then strongly correlated electrons prefer to remain localized on-site.
Thus, whereas LSDA based methods give a good description of U and Np, from Pu onwards additional assumptions or parameters derived from experiment need to be invoked,~\cite{Penicaud,Niklasson,Anisimov_Pu,Shorikov_PuX,Savrasov_Pu_Nature,Svane_AcMetal} diminishing the predictive power of the approach.

The above {\em localization/delocalization} crossover greatly influences the phase diagrams of actinide materials. For example,
the electronic and magnetic properties of U compounds are very different from those of the corresponding Bk compounds.~\cite{Brodsky}
Even for a given actinide element, its various alloys and compounds can
display a wide spectrum of behaviour, from localized to itinerant, due to the large effect that small changes in external or chemical pressure have on the $f$-electron contribution to the chemical bonding. Indeed, Hill~\cite{Hill} suggested that, in actinide compounds, the actinide-actinide distance determines the degree of magnetic order through control of the  $f$-$f$ overlap. However,
many actinide compounds do not follow the systematics based solely on $f$-band formation and it has since become clear that $f$-$d$ and $f$-$p$
hybridizations are equally important in explaining the electronic and magnetic properties of these compounds.

Experimentally, a large number of actinide compounds crystallizes in the NaCl structure. The
mono-chalcogenides AX (X=O,S,Se,Te,Po)  and mono-pnictides (X=N,P,As,Sb,Bi), as well as the actinide mono-carbides,
belong to this class of materials. Experimental studies indicate a trend towards $f$-electron localization
with increasing actinide atomic number and increasing anion size.
For the actinide carbides and nitrides, the large orbital overlap resulting from small anion size competes with the trend towards
more localized $f$-orbitals as the actinide nuclear charge is increased. The resulting competition between band
formation and correlation places these compounds at the borderline of
the localization/delocalization transition. Thus, whether a localized or itinerant $f$-electron model is more adequate to describe
these compounds depends on the details of the underlying electronic structure.

To describe the strongly correlated electrons in the actinide carbides and nitrides,
we use the self-interaction corrected (SIC) local spin density
(LSD) method.~\cite{Svane_SIC,Temmerman_LNP} The SIC-LSD method is an {\it ab initio} approach that corrects for an
unphysical self-interaction of atomic-like localized states in the LSD total energy functional.~\cite{Perdew_Zunger}
The method has previously been applied succesfully to the description of
actinide metals and compounds. Because it is based on total energy considerations, the SIC-LSD methodology enables us to predict the
groundstate valency configuration of the actinide ion and to describe the localization-delocalization transition
that occurs in the carbides and nitrides as the actinide series is traversed.

The balance of this paper is organized as follows.
In section II we give a short description of the SIC-LSD methodology (section IIA) followed by presentation of results for the Uranium compounds (UN and UC) (section IIB), the remaining transuranium compounds (section IIC), and for the localization-delocalization transition (section IID). In section III we present a general discussion of our results in the context of other experimental and theoretical work. Finally, in section IV, we present some concluding remarks.

\section{Electronic structure of actinide pnictides and chalcogenides}
\subsection{SIC-LSD}
The SIC-LSD energy functional, $E^{SIC}$, is obtained from the LSD energy functional,
$E^{LSD}$, by subtracting from it an unphysical self-interaction, $\delta_\alpha^{\mathrm{SIC}}$,
orbital by orbital, for all the occupied orbitals, namely
\begin{equation}\label{eq:sic:1}
E^{\mathrm{SIC}}= E^{\mathrm{LSD}}- \sum_\alpha \delta_\alpha^{\mathrm{SIC}} .
\end{equation}
Since, for the itinerant (delocalized) electrons the self-interaction vanishes, in practice the above sum runs only
over localized orbitals. In the SIC-LSD method both localized and
delocalized states are expanded in the same set of basis functions, and are thus treated on an equal footing.
Different localized/delocalized configurations are realized by assuming different numbers and combinations of localized
states - here $f$-states on actinide-atom (A) sites. Since the different localization scenarios constitute distinct
local minima of the same energy functional, $E^{SIC}$, their total energies may be compared and
the global energy minimum then defines the ground state total energy {\em and} the valence configuration
of the A-ion. This latter is defined as the integer number of electrons available for band formation, namely
$N_{val}=Z-N_{core}-N_{SIC}$
%\begin{equation}
%N_{val}=Z-N_{core}-N_{SIC},
%\end{equation}
where Z is the atomic number, $N_{core}$ is the number of core (and semicore) electrons, and $N_{SIC}$ is the number of localized, i.e. self-interaction corrected, states.
We will use either the $f^n$ or the A$^{m+}$ nomenclature to describe the actinide-ion configuration, implying
$n=N_{SIC}$ and $m=N_{val}$, respectively. Note that the number of $f$-electrons on a given ion may be larger than $n$, since, in addition to the localized
$f$-states, the band states contribute to the total $f$-electron count.

The SIC-LSD approach has been implemented using the tight-binding linear muffin-tin orbital (LMTO) method in
the atomic sphere approximation (ASA).~\cite{Andersen_LMTO_ASA} The spin-orbit interaction has been explicitly added to the scalar-relativistic
one-particle Hamiltonian, and included in the self-consistency cycle. All the compounds investigated in the
present paper have been shown experimentally to crystallize in the NaCl structure. In order to
improve the packing, empty spheres have been introduced on high-symmetry interstitial sites. Two uncoupled
energy panels have been considered when constructing the LMTO's, with $s$, $p$, $d$, and $f$ orbitals on all spheres.
The valence panel includes the 7$s$, 6$d$ and 5$f$ orbitals
on the actinide atom, and the 2$s$ and 2$p$ orbitals on the N and C atoms, with the remaining orbitals downfolded.~\cite{Lambrecht_downf}
The semicore panel comprises the actinide 6$p$ states, all other channels being downfolded.

%The relative ratio of the spheres used is 2.78, 1.84, 1.22 for
%the actinide, anion, and the empty sphere respectively.
\begin{figure}
\begin{center}
\includegraphics[width=120mm,clip,angle=-90]{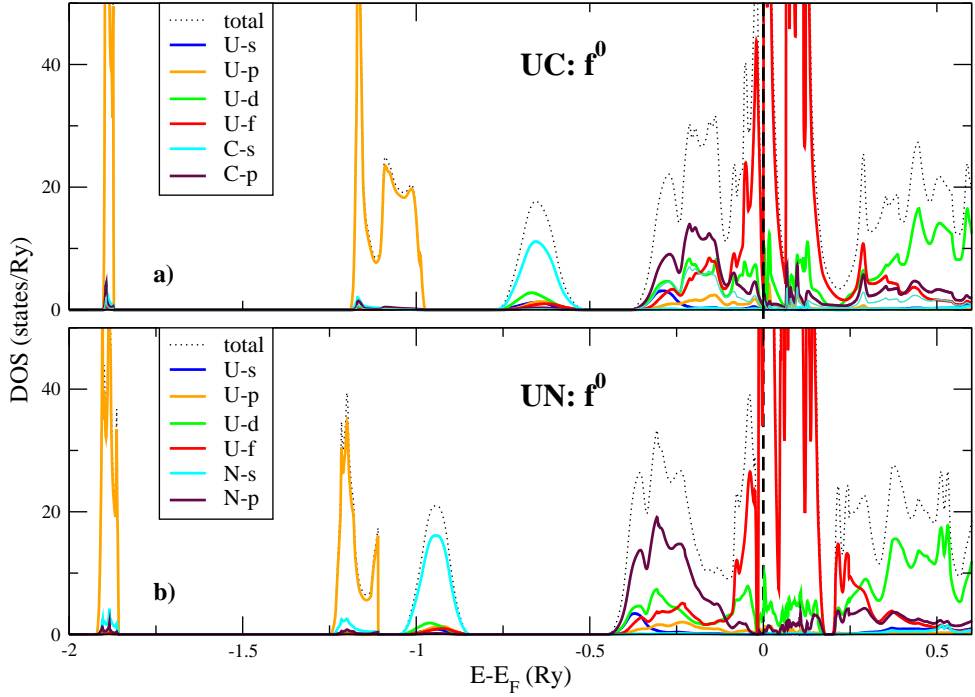}
\caption{
\label{UCN}
Density of states of a) UC, and b) UN in the LSD approximation.
}
\end{center}
\end{figure}

\subsection{UN and UC}
Given the large extent of the U 5$f$-orbitals, combined with the small radii of the carbide and nitride atoms,
UC and UN have always been assumed to qualify for a standard bandstructure description, with the $f$-electrons
treated as itinerant. Correspondingly, the electronic structure of these compounds has been studied using various
bandstructure approaches.~\cite{Weinberger_UN, Mallet,Brooks_UNCO,Evarestov,Weck,Hasegawa_UC,Trygg_UC,Atta-Fynn}
The resulting values for the lattice parameters of both UC and UN, as well as the elastic
constants for UC~\cite{Trygg_UC} have been found to be in good agreement with experiment.

In the SIC-LSD approach, the LSDA band picture is reproduced when all the $f$-electrons
are treated as delocalized (U($f^0$) $\equiv$ U$^{6+}$ configuration).
The densities of states (DOS) corresponding to this configuration are shown in figures \ref{UCN}a and \ref{UCN}b for UC and UN respectively. The two DOS are overall very similar, and in agreement with earlier LSDA based calculations, for example
by Trygg {\it et al.} for UC,~\cite{Trygg_UC} and by Samsel-Czekala {\it et al.} for UN,~\cite{Samsel} where the different
features are explained in
detail. The electronic structure is characterized by the overlap of the U-$f$ states with the C/N $p$-states and
the large narrow U-$f$ peak situated at the Fermi level.

\begin{figure}
\begin{center}
\includegraphics[width=120mm,clip,angle=-90]{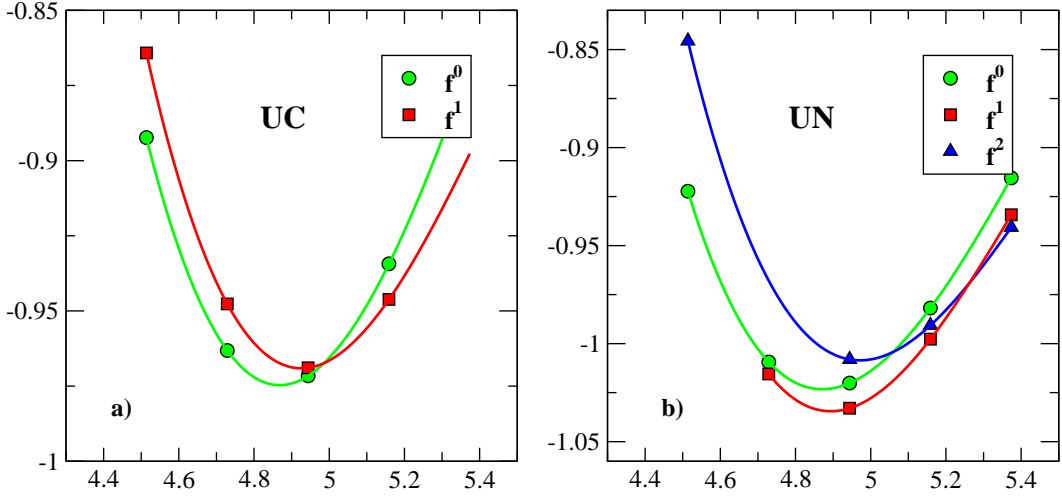}
\caption{
\label{UCE}
Total energy as a function of lattice parameter
for a) UC, and b) UN for different localization/delocalization configurations.
}
\end{center}
\end{figure}

The results of our total energy minimization
as a function of lattice parameter and different localized/delocalized$f$-electron
configurations (including the fully delocalized $f^0$ configuration)
are shown in figures \ref{UCE}a and \ref{UCE}b. For UC, in Fig. \ref{UCE}a, we find the $f^0$ configuration to be energetically
most favourable,
confirming that the LSDA based DOS in Fig. \ref{UCN}a adquately represents the corresponding
electronic structure. For UN,
we find the global energy minimum to occur in the $f^1$ configuration, as can be seen in Fig. \ref{UCE}b,
indicating that
the LSD picture of Fig. \ref{UCN}b might not be a good representation of the UN groundstate as $f$-electron localization
is starting to set in.
The reason for the difference in localization behaviour can
be traced to the fact that the nitride is more electronegative than the carbide.
In the DOS (Fig. \ref{UCN}),
compared to the rather considerable $p$-$f$ overlap in UC, we observe that for UN
the $p$ band is situated lower in energy
with respect to the Fermi level and a valley develops between the $p$ and $f$-states.
With the reduced $p$-$f$ overlap in UN, hybridization becomes less predominant,
the gain in band formation energy is reduced, and the gain in localization energy becomes relatively more important.

There is experimental evidence that supports the picture of increased localization in UN compared to UC.
For UC, the calculated Fermi surface in the LSDA agrees well with the measured de Haas-van Alphen frequencies,~\cite{Hasegawa_UC} indicating that the
5$f$-electrons are indeed delocalized. The XPS and BIS measurements on UC~\cite{Ejima}
confirm this itinerant behaviour, but the observation of 4$f$-satellites in the spectra indicates that correlations among
the 5$f$ electrons already have a noticeable effect.
The interpretation of experimental results is less straightforward for UN, where it remains unclear whether a localized, delocalized,
or dual localized/delocalized picture can best account for the observed properties.~\cite{Samsel,Norton,Solontsov}
ARPES studies on UN seem to reveal some degree of localization of the U $f$-states, with two non-dispersive bands detected
in the vicinity of the Fermi level,~\cite{Ito_UN} compared to a single dispersive band at the Fermi level in UC.~\cite{Ito_UC}
Specific heat measurements of respectively $\gamma$=18.7 mJ/K$^2$mol for UC, and $\gamma$=49.6 mJ/K$^2$mol for UN,
are a clear indication of considerable renormalization of the $f$-bands by
the electron-electron interactions not accounted for in the LSDA.
\begin{figure}
\begin{center}
\includegraphics[width=120mm,clip,angle=0]{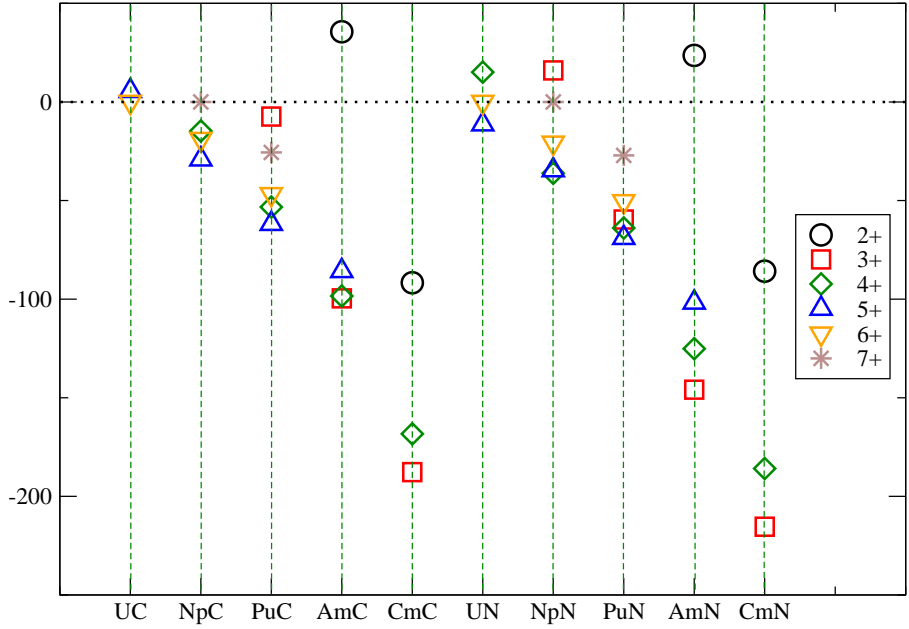}
\caption{
\label{energies}
SIC-LSD total energies for the actinide nitrides and carbides from U to C. For each compound the energies for a number
of different valency configurations are given (in mRy/formula unit) relative to the LSDA total energy.
}
\end{center}
\end{figure}

Overall, experiment clearly indicates a growing impact of correlations from UC to UN, but it is less obvious whether,
combined with a narrowing of the $f$-peak at the Fermi level, an actual $f$-electron localization
transition will occur. For UN the SIC-LSD calculations do indicate such a transition, however it should be noted that
the calculated energy difference, E$_{f^0}$-E$_{f^1}$, between the $f^0$ and $f^1$ configurations is very small. Indeed, this is true both for UC and UN where the respective energy differences are -5 mRy and +10 mRy. This approximate
degeneracy of configurations implies that the gain in energy from band-formation
and the gain in SIC energy from localization are of similar magnitude indicating an electronic structure at the border
of the localization-delocalization transition. Rather than a fully developed itinerant-$f$ groundstate for UC and
a localized $f^1$ groundstate for UN, the electronic structure is intermediate between $f^0$ and $f^1$, with the former
configuration having relatively
more weight in UC and the latter configuration having relatively more weight in UN. In both the $f^0$ and $f^1$ cases, a significant amount
of itinerant $f$-electrons is found (the total $f$-electron count is $\sim$ 3.0 for UN and $\sim$ 2.8 for UC) and the LSD description
is acceptable. The observed difference in total energies
correctly reproduces the increasing influence of electron-electron correlations from UC to UN. However the actual electronic structure
of these compounds is more complex than can be reproduced by either the localized or itinerant $f$ limit
of the SIC-LSD single Slater determinant wavefunction,
and would need to be addressed by more sophisticated approaches.\cite{Georges,Lueders_LSIC}

\subsection{Transuranium nitrides and carbides}

Applying SIC-LSD to the carbides and nitrides beyond U, we observe a trend towards increasing localization with increasing actinide
atomic number. The results of our total energy calculations are summarized in figure \ref{energies}.
Here, for each of the actinide compounds AC (A=U, Np, Pu, Am, and Cm) and AN, the calculated total energies
for a number of configurations are given relative to the LSD total energy. Positive and negative energies respectively
indicate configurations that are less or more
favourable than the fully delocalized scenario. Only for UC is the LSD configuration the groundstate, as was discussed earlier.
From NpC to CmC, the groundstate valency configuration gradually changes from A$^{5+}$ to A$^{3+}$. This trend repeats itself from UN to CmN.
As expected the increase in nuclear charge leads to the contraction of the 5$f$-orbitals, and thereby a decrease
in overlap with neighbouring sites. The localization energy becomes relatively more important than the band formation energy,
leading gradually to the localization of an increasing number of $f$-electrons.
A detailed look at the total energies in Fig. \ref{energies} reveals that, on average, for a given actinide ion,
the nitride displays a lower groundstate valency than the corresponding carbide, as is also shown in Table \ref{groundstate}.
Most importantly, a clearly preferred groundstate configuration (a configuration that has a substantially lower energy than the rest) emerges in the nitride series
for AmN and CmN, for the carbide series such a configuration emerges only for CmC.
The electronic structure of the early actinide compounds beyond UN and UC thus remains a complex manifold of different valency
configurations, closely separated in energy, with the contribution from the more localized configurations becoming gradually more important relative
to the less localized configurations as we move through the actinide series.
\begin{figure}
\begin{center}
\includegraphics[width=120mm,clip,angle=-90]{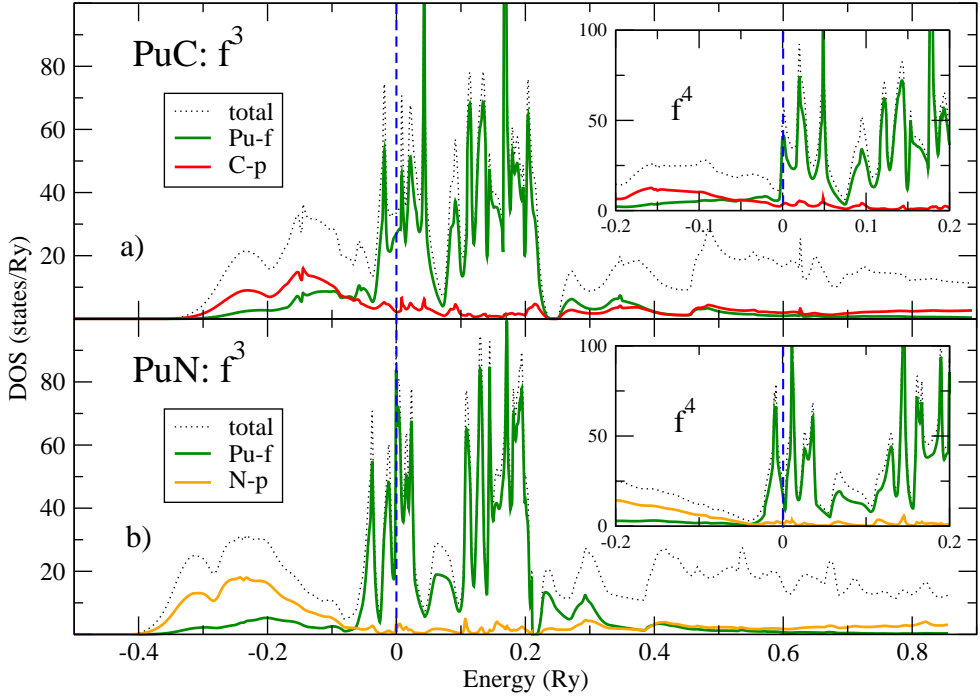}
\caption{
\label{UCP}
Density of states for a) PuC in the $f^3$ configuration, and $f^4$ configuration (inset),
b) PuN in the $f^3$ configuration, and $f^4$ configuration (inset).
}
\end{center}
\end{figure}

We can try to understand this configuration degeneracy on the basis of the DOS for PuC and PuN in figure \ref{UCP}.
Notice that, in these plots, only the band states are displayed, i.e. the itinerant valence states including the delocalized
$f$-states. Specifically, the localized $f$'s are not shown since the SIC-LSD approach, which, after all, is a one-electron ground-state theory, does not
give accurate removal energies of localized states due to electron-electron interaction (multiplet) effects,\cite{Svane_DMFT} and
the neglect of screening and relaxation effects.\cite{Temmerman_Pr}
In the SIC-LSD, if a given $f$-state hybridizes strongly with the broad $p$-bands the gain in band formation
energy can overcome the gain in localization energy. If, on the other hand, the $f$-states are restricted to
a narrow band the gain in band formation energy is small and localization is energetically more favourable.
%bforms a narrow band, rather than
%hybridizing strongly with the broad anion $p$-band, the localization energy tends to outcompete the band formation energy.
In the $f^3$ configuration
of Fig. \ref{UCP}, for both PuC and PuN, the itinerant $f$-states fill a narrow peak below the Fermi level which is, however,  still strongly hybridized  with the broad $p$-band.
Localizing an additional $f$-state leads to the $f^4$ configuration depicted in the respective insets to Fig. \ref{UCP}, 
with an associated gain in localization (i.e. self-interaction correction) energy. However,
the Fermi level has also moved closer to the $p$-band resulting in a depopulation of Fermi energy $f$-states
and a considerable loss of hybridization energy.
Whilst the gain in energy from localizing a given $f$-state is approximately the same for both PuC and PuN, the corresponding
loss in band formation energy differs. In PuC $f-p$ hybridization is more pronounced than in PuN where
the $p$-states are further separated from the $f$-states due to the increased electronegativity.
In PuN, with a calculated total energy difference E$_{f^3}$-E$_{f^4}$=+5 mRy, both configurations contribute almost
equally to the groundstate electronic structure. In PuC, the $f^3$ configuration, with one more itinerant $f$-state,
is slightly more favourable (E$_{f^3}$-E$_{f^4}$=-9 mRy) due to the increased hybridization.
The total (localized+itinerant) $f$-electron
count for the different scenarios shown in Fig. \ref{UCP} is
PuC($f^3$)=5.17, PuC($f^4$)=5.15, PuN($f^3$)=5.38, and PuN($f^4$)=5.35. Clearly, for both compounds the total number
of $f$-electrons changes only marginally through the $f^3$ $\rightarrow$ $f^4$ transition, however there are overall 0.2 $f$-electrons less in PuC than in PuN  due to the increased
orbital overlap.

Over all then, the bonding in PuN is more ionic in nature compared to the more covalent bonding in PuC. This difference is even more
noticeable if we try to localize a fifth $f$-electron. In this case, we find the resulting Pu$^{3+}$ configuration to be almost degenerate
with both the Pu$^{4+}$($f^4$), and the Pu$^{5+}$($f^3$) configurations in PuN, whereas in PuC it is energetically unfavourable by
$\sim$ 50 mRy. From the inset of Fig. \ref{UCP}a, we see in the $f^4$ configuration of PuN a narrow $f$-peak
just
below the Fermi level, a level that can then be localized with a moderate loss in band formation energy.
On the other hand, in the $f^4$
configuration of PuC (inset of Fig. \ref{UCP}b) the Fermi level is
below the first peak in the $f$-states. As a consequence localizing an additional $f$-state would imply
depopulating the C $p$-band which would then result in charge transfer. Clearly,
the associated loss in Madelung and hybridization energy would then be considerably larger than the
gain in localization energy and is, therefore, unfavourable.

Again, we are driven to assume that the actual electronic groundstate configuration of these early actinide carbides and nitrides
is more complex than any of the limiting configurations in the almost degenerate manifold can describe.
If the true groundstate is fluctuating between a given set of configurations, as is the case for PuC (where the $f^3$ and $f^4$
configurations are close to being energetically equivalent), one would expect some intermediate degree of localization. With
the fully delocalized LSD ($f^0$) configuration and the fully localized $f^5$ configuration being energetically rather unfavourable
compared to the $f^3$ configuration, by respectively 61 mRy and 53 mRy, these configurations should not
contribute significantly to the electronic structure. Thus it is reasonable to assume that the electronic structure of PuC consists of a manifold of coexisting
localized and delocalized $f$-states, similar to the two main components $f^3$ and $f^4$. The fact that $f^3$ gives the global energy
minimum, and that localizing an additional $f$-electron is only slightly less favourable, might also indicate an intermediate scenario
with three localized $f$-electrons and a strongly renormalized itinerant $f$-peak.

It is interesting to note that for compounds at the end of actinide series, the trivalent ($f^5$) groundstate configuration becomes noticeably more favourable than both the
tetravalent ($f^4$) and pentavalent ($f^3$) configurations. In the carbide series this clearly preferred groundstate configuration
emerges only for CmC, whereas in the nitride series
the multi-configuration degeneracy is already lifted for AmN.
These observations can then be taken as a clear indication that the trend towards
localization is more pronounced for the nitride compounds than for the carbides.

\begin{table}
\begin{ruledtabular}
\begin{tabular}{|c|c|c|c|c|}
   & Ground state & alat$_{LSD}$ & alat$_{calc}$ ({\AA}) & alat$_{exp}$ ({\AA}) \\
\hline
  UC       & f$^0$ (U$^{6+}$)& 4.86 & 4.86 & 4.960  \\
\hline
  NpC      & f$^2$ (Np$^{5+}$)& 4.85 & 4.93  & 4.999 \\
\hline
  PuC      & f$^3$ (Pu$^{5+}$)& 4.86 & 4.93  & 4.965 \\
\hline
  AmC      & f$^5$/f$^6$ (Am$^{4+}$/Am$^{3+}$)& 4.87  & 4.97/5.04  & -  \\
\hline
  CmC      & f$^7$ (Cm$^{3+}$)& 4.86  & 4.99 & -\\
\hline
\hline
  UN       & f$^1$ (U$^{5+}$) & 4.87  & 4.89  & 4.890  \\
\hline
  NpN      & f$^2$/f$^3$ (Np$^{5+}$/Np$^{4+}$) & 4.87  & 4.90/4.98 & 4.897  \\
\hline
  PuN      & f$^3$ (Pu$^{5+}$) & 4.89 & 4.93  & 4.905  \\
\hline
  AmN      & f$^6$ (Am$^{3+}$)  & 4.92  & 5.03  & 4.995   \\
\hline
  CmN      & f$^7$ (Cm$^{3+}$)  & 4.90  & 5.02  & 5.041 \\
\hline
\end{tabular}
\end{ruledtabular}
\caption{Actinide carbide/nitride data: Column 2, groundstate configuration. Column 3, calculated lattice parameter in the
LSD approximation.
Column 4, calculated lattice parameter in the groundstate configuration. Column 5, experimental lattice parameter.\cite{Rossat,Brodsky,Benedict_PuC}}
\label{groundstate}
\end{table}

In Table \ref{groundstate} the calculated lattice parameters in the groundstate configuration of the actinide carbides
and nitrides (column 4) are compared to the corresponding experimentally observed values (column 5).
The agreement is generally very good, with a deviation from experiment of around 1.5 \% for the carbides and 0.5 \% for
the nitrides.
However, at odds with experiment, the calculations predict the lattice parameters of a particular actinide carbide and nitride to be approximately equal. In fact the measured lattice parameters of the carbides (UC, NpC, and PuC)
tend to be larger than those of their nitride counterparts by almost 2\% (see Table \ref{groundstate}).
Given the more covalent nature of bonding in the carbides, one would actually expect the opposite to happen,
i.e. that the nitride lattice parameters would be relatively larger; although this effect may be somewhat counterbalanced by the
increased overlap due to the slightly smaller nitride anion.

On the experimental side it turns out that actinide monocarbides exist as defect structures, AC$_{1-x}$.\cite{Erdos} Furthermore, the fraction of randomly distributed
C vacancies
affects the measured lattice parameters. For PuC$_{1-x}$ a range of different lattice parameters is found depending on {\it x} as well as temperature
(the value quoted in Table \ref{groundstate} is for $x \sim 0.2$ and $T \sim 100 K$).\cite{Benedict_PuC}
Using a supercell consisting of four PuC formula units with
one single carbon atom removed, we studied the effect of C vacancies on the total energy and lattice parameter.
We find the groundstate of PuC$_{0.75}$ to remain Pu$^{5+}$ with 3 $f$-electrons localized on each Pu site as in PuC, but the
equilibrium lattice parameter is now increased to 4.99 Angstrom. Thus, it appears that the relatively large lattice parameters
of the actinide carbides, compared to the nitrides, are related to the presence of C vacancies in the actual compounds, rather than to differences
in the electronic structure between PuC and PuN.
No experimental values for AmC and CmC seem to exist.

\begin{figure}
\begin{center}
\includegraphics[width=120mm,clip,angle=0]{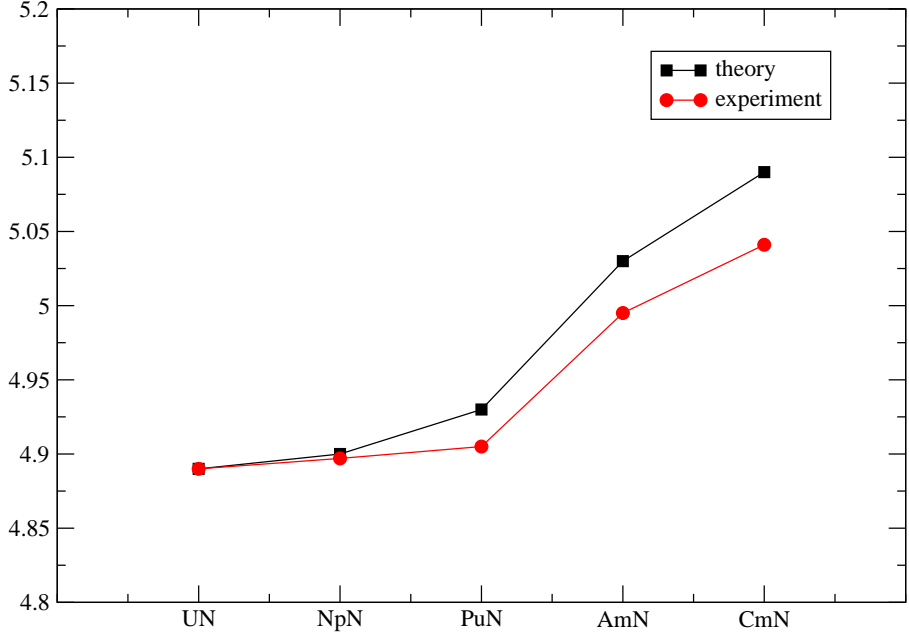}
\caption{
\label{alat}
Lattice parameters (in \AA) of the actinide nitrides: experiment versus theory.
}
\end{center}
\end{figure}

\subsection{Localization-delocalization transition in actinide nitrides}
%In figure \ref{alat} we compare the experimental and theoretical lattice parameters of
%the actinide nitrides from UN to CmN. Most noticeable is the sudden increase that is
%observed from PuN to AmN, and which is well reproduced by the calculated values.
From the total
energies in figure \ref{energies} we observe that in the nitride series a preferred groundstate configuration
only emerges at AmN and CmN. This sudden localization that occurs between PuN and AmN is reflected in the jump
in the corresponding lattice parameters that is seen in Fig. \ref{alat}. The abrupt increase in the measured lattice
parameters is well reproduced by our calculated values; in the SIC-LSD calculations it is clearly associated with the fact that
the localized $f$-states no longer participate in bonding.
An increasing lattice parameter from UN to AmN is also
observed in our LSD calculations (column 3 of Table \ref{groundstate}), as well as in earlier LSDA or GGA based
calculations.\cite{Brooks_UNCO,Atta-Fynn} However,
in the case of standard LSD and GGA calculations, the change in lattice parameter is gradual and a result of 
gradual narrowing of $f$-bands with increasing actinide atomic number; as distinct from a localization transition as in SIC-LSD. The jump between
PuN and AmN has not been observed in previous LSD and GGA calculations.

\begin{figure}
\begin{center}
\includegraphics[width=120mm,clip,angle=-90]{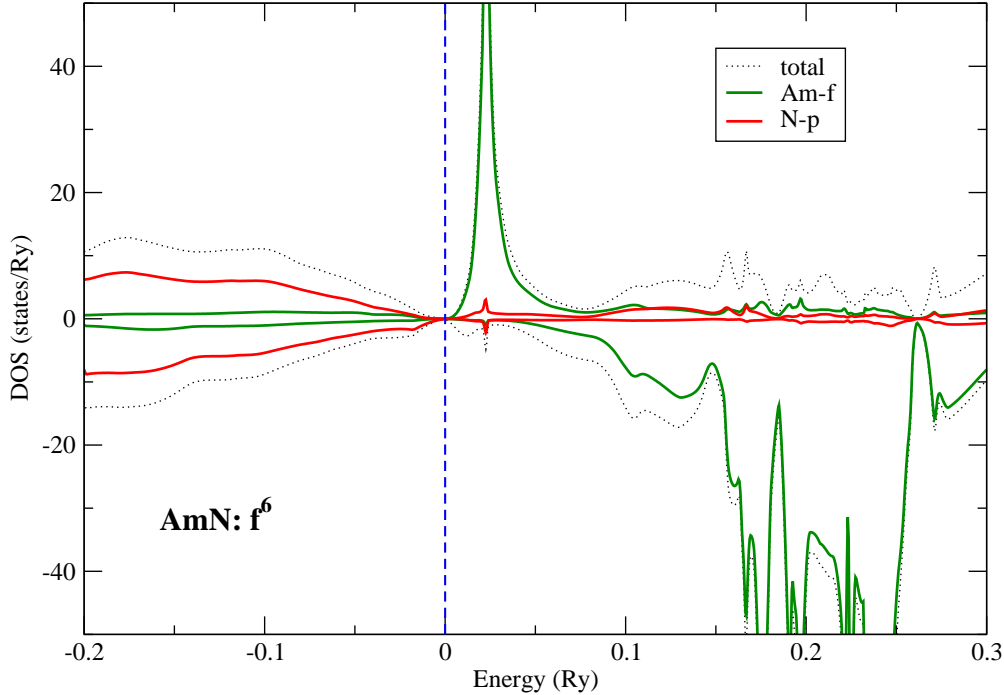}
\caption{
\label{amn}
DOS of AmN in the f$^6$ groundstate configuration.
}
\end{center}
\end{figure}

From our SIC-LSD calculations we thus predict a delocalization-localization transition to occur between PuN and AmN and that
the resulting groundstate configuration for AmN is Am$^{3+}$ with 6 localized $f$-electrons, the DOS of which is shown in figure \ref{amn}.
With the N-$p$ band capable of accepting three valence electrons through charge transfer and hybridization, in the trivalent groundstate
the $p$-band is completely filled, and the Fermi level is situated below the empty $f$-peak. AmN is apparently close to being a semiconductor.
As we noted in an earlier publication on the Americium pnictides,\cite{Petit_AmX} the vanishingly small DOS at the Fermi level of AmN is not in agreement
with the rather high values for the observed temperature independent paramagnetic susceptibility.\cite{Kanellakopulos}
However, a subsequent photoemission study by Gouder {\it et al.}\cite{Gouder_AmX}
seems to have confirmed both the predicted 5$f^6$ groundstate as well as the semiconducting character of AmN. In a separate publication,
the experimentally observed high value for the susceptibility was explained by a
Van Vleck mechanism.\cite{Ghosh_AmX}

\section{General discussion}
The strength of the SIC-LSD is that it is a parameter free theory of the ground state which allows one to determine the valence state by comparing energies. Furthermore, it casts light on the systematics of localization-delocalization transitions in that the self-interaction is non-zero for localized states but vanishes for band states; the latter being the reason 
 why the LSDA remains valid for the itinerant $f$-states. Given that SIC-LSD is a ground state theory it relates mainly to occupied states.  Accordingly
in the calculated DOS, the delocalized $f$-states tend to be situated at too low binding energies due to the fact that the correlations among the electrons occupying these
states are not fully accounted for.  Thus, for the DOS in Fig. \ref{amn}, even though it correctly reproduces the pseudogap,  the DOS cannot straightforwardly be mapped
onto the main features of the XPS and BIS measurements. Indeed, as calculated, the positions of both the localized and delocalized $f$-states are not well defined.
In the LDA+U approximation, guided by experiment, an effective U parameter is introduced that separates the $f$-manifold into
the lower and upper
Hubbard bands and removes the $f$-degrees of freedom from the Fermi level.\cite{Anisimov}
In a study of AmN, assuming localized $f$-states, and using a U parameter of 2.5 eV, Ghosh {\it et al.}\cite{Ghosh_AmX} were
able to reproduce the 5$f$ binding energy positions of the
photoemission measurements ($\sim$ 2.5 eV).
As indicated above, in the fully first-principles SIC-LSD no parameter for modeling the strong correlations is introduced. However,
we can attempt to estimate the position of the 5$f$ states from our calculated values for respectively the SIC
corrected $f$-states and the band $f$-states by using a transition state
argument.\cite{Janak,Svane_GaAs} For AmN, this gives a binding energy of 3.3 eV,
which is about 30 \% larger than the value of 2.5 eV determined from experiment.

The LDA+U approach has similarly been applied to PuN\cite{Shorikov_PuX} and PuC.\cite{Gouder_PuC} In both cases the localized $f$-manifold consists of five $f$-electrons, as a result
of which no itinerant $f$-states appear at the Fermi level of the corresponding DOS. This is in contradiction to the results from photoelectron emission studies
where a triplet of 5$f$ related features is observed, including a strong peak at the Fermi level.\cite{Havela_PuN,Gouder_PuC} The SIC-LSD calculations predict an $f^3$ groundstate
configuration, albeit with strong contributions from the $f^4$, and in the case of PuN, also the $f^5$ initial state configurations. Both the $f^3$
and the $f^4$ configurations are characterized by coexisting localized and delocalized $f$-states, which, as can be seen from Fig. \ref{UCP},
results in a large DOS at the Fermi level, in agreement with photoemission experiments. For completeness, we should  mention that the photoelectron
measurements by Havela {\it et al.}\cite{Havela_PuN} at the time were interpreted in terms of a 5$f^3$ groundstate for PuN, in good agreement with our predicted groundstate
configuration.\cite{Petit_PuX}

Finally, with respect to the high temperature behaviour of the actinide nitrides and carbides, it has been observed that for the nitrides the thermal
conductivity decreases from UN to PuN and it was suggested that this decrease is caused by a decrease in the electronic contribution to the thermal
conductivity.\cite{Arai} This explanation would thus agree with the observed tendency towards a decreasing number of itinerant electrons with increasing actinide atomic number.
An interesting feature in this context is the variation of the lattice parameter with temperature observed experimentally; a dependency that has been previously modeled by
molecular dynamics simulations.\cite{Kurosaki}
In our SIC-LSD calculations, an increasing lattice parameter would result in the more localized scenarios becoming gradually more
favourable. This would imply, that as far as the $f$-electron contribution is concerned, the thermal conductivity for a given compound would decrease with increasing
temperature. The measured data\cite{Arai} do seem to show such a trend at least for UN and NpN. Here however, it should be noted that
SIC-LSD
is a ground-state theory that does not take into account all possible fluctuation effects that are associated with increasing temperature.
An extension of the present work to include finite temperature effects is currently being pursued based on the local self-interaction correction implemented
in the multiple scattering theory.\cite{Lueders_LSIC} In combination with the coherent potential approximation and the disordered local moments theory,
this approach allows one to study possible spin and valence fluctuations as a function of temperature. 
The utility of this methodolgy has already been demonstrated for Ce,\cite{Lueders_LSIC}
and Gd.\cite{Hughes}

\section{Conclusion}
We have studied the groundstate electronic structure of the actinide mono-nitrides and mono-carbides.
A trend towards increased $f$-electron localization as a function of actinide atomic number has been observed, which is slightly
more predominant in the nitrides than in the carbides. With the exception of UC, the light actinide compounds are best described
in terms of a manifold of several coexisting localized/delocalized configurations. A localization transition occurs in the late actinides which 
results in a jump in lattice parameter from PuN to AmN. The valence electron manifold of all these compounds
reproduces the main features of the photoemission experiments, including the band gap (pseudo gap) that is observed in AmN.

This research used resources of the Danish Center for Scientific Computing (DCSC) and of the National Energy Research Scientific
Computing Center (NERSC). Research supported in part (GMS) by the Division of Materials Science and Engineering, Office of Basic Energy Science, U.S. Department of Energy.
%\bibliography{worldnuclear}http://www.world-nuclear.org
%\bibliography{/Users/leonpetit/desktop/lp/ARTICLES/BIBTE/leon_ref}
%\bibliography{accarb}

\end{document}